\documentclass[aps, prd, preprint, superscriptaddress, tightenlines, nofootinbib]{revtex4}

\pdfoutput=1

\usepackage[english]{babel}
\usepackage{bbm} 
\usepackage{bm} 
\usepackage{array}
\usepackage{amsthm}
\usepackage{amssymb}
\usepackage{amsmath}
\usepackage{graphicx}
\usepackage{color}
\usepackage{verbatim}
\newcommand{\be}{\begin{eqnarray}}
\newcommand{\ee}{\end{eqnarray}}

\renewcommand{\d}{\mbox{${\rm d}$}} 
\newcommand{\lp}{\ell_{\rm p}}
\newcommand{\mpl}{m_{\rm p}}
\newcommand{\gn}{G_{\rm N}}

\newcommand{\Rh}{R_{\rm H}}

\usepackage{datetime}
\settimeformat{ampmtime}

\usepackage{float}

\usepackage[bookmarks,breaklinks,plainpages=false,pdfpagelabels]{hyperref} 
	\hypersetup{linkcolor=red,citecolor=blue,filecolor=dullmagenta,urlcolor=darkblue} 

\usepackage{soul}

\usepackage{slashbox}

\definecolor{grey}{rgb}{0.4,0.4,0.4}
\definecolor{dullmagenta}{rgb}{0.4,0,0.4}
\definecolor{darkblue}{rgb}{0,0,0.4}
\definecolor{midblue}{rgb}{0,0,0.5}
\definecolor{midred}{rgb}{0.5,0,0}
\definecolor{orange}{rgb}{1,0.5,0}
\definecolor{lightbrown}{rgb}{0.75,0.5,0.25}
\definecolor{tan}{cmyk}{0.14,0.42,0.56,0}
\definecolor{djunglegreen}{cmyk}{0.99,0,0.52,0}
\definecolor{lightgreen}{rgb}{0,1,0}
\definecolor{olivegreen}{cmyk}{0.64,0,0.95,0.40}
\definecolor{midgreen}{rgb}{0.0,0.675,0.0}
\definecolor{darkgreen}{rgb}{0,0.5,0}

\newcommand{\qq}{\qquad}

\newcommand{\vs}{\vspace}

\renewcommand{\.}{\hspace{0.5mm}}

\newcommand{\Crm}{\ensuremath{\mathrm{C}}}

\newcommand{\Nrm}{\ensuremath{\mathrm{N}}}

\newcommand{\Prm}{\ensuremath{\mathrm{P}}}

\newcommand{\erm}{\ensuremath{\mathrm{e}}}

\newcommand{\Ccal}{\ensuremath{\mathcal{C}}}

\newcommand{\Scal}{\ensuremath{\mathcal{S}}}

\newcommand{\Nbbm}{\ensuremath{\mathbbm{N}}}

\newcommand{\Rbbm}{\ensuremath{\mathbbm{R}}}

\newcommand{\lbm}{\ensuremath{\bm{l}}}

\newcommand{\defas}{\mathrel{\mathop :}=} 

\renewcommand{\d}{\ensuremath{\mathrm{d}}}

\newcommand{\eg}{e.g.}
\newcommand{\ie}{i.e.}

\newcommand{\cf}{c.f.}

\begin{document}

\title{\bf Consistent Cosmic Microwave Background Spectra\\
from Quantum Depletion}

\author{Roberto Casadio}
\email{casadio@bo.infn.it}
\affiliation{Dipartimento di Fisica e Astronomia,
	Alma Mater Universit{\`a} di Bologna,
	via Irnerio~46,
	40126 Bologna,
	Italy}
\affiliation{I.N.F.N.,
	Sezione di Bologna,
	via B.~Pichat~6/2,
	40127 Bologna,
	Italy}

\author{Florian K{\"u}hnel}
\email{florian.kuhnel@fysik.su.se}
\affiliation{The Oskar Klein Centre for Cosmoparticle Physics,
	Department of Physics,
	Stockholm University,
	AlbaNova,
	106\.91 Stockholm,
	Sweden}

\author{Alessio Orlandi}
\email{aorlandi@bo.infn.it}
\affiliation{Dipartimento di Fisica e Astronomia,
	Alma Mater Universit{\`a} di Bologna,
	via Irnerio~46,
	40126 Bologna,
	Italy}
\affiliation{I.N.F.N.,
	Sezione di Bologna,
	via B.~Pichat~6/2,
	40127 Bologna,
	Italy}

\date{\formatdate{\day}{\month}{\year}, \currenttime}

\begin{abstract}
Following a new quantum cosmological model proposed by Dvali and Gomez,
we quantitatively investigate possible modifications to the Hubble parameter
and following corrections to the cosmic microwave background spectrum.
In this model, scalar and tensor perturbations are generated by the quantum
depletion of the background inflaton and graviton condensate respectively.
We show how the inflaton mass affects the power spectra and the tensor-to-scalar
ratio. Masses approaching the Planck scale would lead to strong deviations, 
while standard spectra are recovered for an inflaton mass much smaller than
the Planck mass.
\end{abstract}

\maketitle

\section{Introduction}
\label{sec:Introduction}
\setcounter{equation}{0}

\noindent
Cosmology is a well understood branch of physics today.
The data gathered since the early 1990s has allowed us to study
thoroughly the origin of the observable Universe.
One of the most important theoretical achievements is the
inflationary scenario\footnote{Not feeling the need to
explain the importance of inflationary models, we just mention
two of the many review papers that have been written on the subject,
namely Baumann~\cite{Baumann:2009ds} and Linde~\cite{Linde:2007fr}.}.
Cosmological inflation was introduced by Starobinski~\cite{Starobinski}
and Guth~\cite{Guth:1980zm} in the early 1980s in order to explain why
the observable Universe is so homogeneous and flat.
It has been known since 1929~\cite{Hubble:1929ig} that the Universe is
expanding, but if we trace this expansion backward in time, we see that
the present-day sky was made of many causally disconnected patches.
This either means that the homogeneity we observe today cannot have been
produced by any interaction in the past,
or, one can argue that the Universe was actually smaller than we expect
from our simple extrapolation and it therefore underwent some
kind of accelerated expansion in a very short time just after its creation.
This brief expansion is called cosmological inflation,
and is generally assumed to have occurred from about $10^{-36}$ to
$10^{-32}\,$seconds after the (presumed) origin of the Universe.
\par
The large amount of data collected by experiments such as COBE,
BOOMERanG, WMAP and Planck strongly reassures us of the validity
of the so called $\Lambda$CDM cosmological model.
Nevertheless, we observe that at large angular scales the Universe
displays a little less anisotropy than we expect~\cite{lowmultipole}.
In this paper we analyse quantitatively the corpuscular model proposed
by Dvali and Gomez in Ref.~\cite{Dvali:2013eja,Dvali:2014gua}, and
show that corrections to the standard power spectrum are larger
at large scales, and that the tensor-to-scalar ratio is naturally
suppressed by cumulative quantum-depletion effects
for inflaton masses approaching the Planck scale.
Although conceptually very different from the standard inflationary
scenario, the model leads to predictions that satisfy current experimental bounds.
In particular, for inflaton masses much smaller than the Planck scale
(\eg~around $10^{13}\,$GeV),
corrections to the standard power spectrum
are well within cosmic variance around the standard predictions.
\par
We will briefly review the original idea of
self-gravitating Boson condensates to describe black holes in Sec.~\ref{sec:The-Graviton-Condensate}, 
and show how this model can be applied to inflation in Sec.~\ref{sec:Quantum-Inflationary-Cosmology}.
In Sec.~\ref{sec:Cosmological-Parameters}, we numerically
solve the time-evolution equation and show how various cosmological
parameters behave as functions of time, and how the cosmic microwave background (CMB)
power-spectrum is affected.
Finally, we will comment about our results and outline possible further developments
in Sec.~\ref{sec:Discussion-and-Outlook}.

We will mostly use units such that $c = 1$, $\hslash = \mpl\, \lp$,
and the Newton constant $G_{\Nrm} = {\lp}/{\mpl}$, where $\lp$ and $\mpl$
are respectively the Planck length and Planck mass, except for
Section~\ref{sec:Cosmological-Parameters}, where we shall also use
astronomical units.

\section{The Graviton Condensate}
\label{sec:The-Graviton-Condensate}
\setcounter{equation}{0}

\noindent
In recent years, a new way of describing the quantum aspects of black holes
was proposed in Refs.~\cite{DvaliGomez}.
The idea is simple and straightforward:
black holes can be seen as a self-gravitating Bose-Einstein condensates (BECs)
of gravitons at a critical point, with Bogoliubov modes that become degenerate
and nearly gapless, representing the holographic quantum degrees of freedom
responsible for the black-hole entropy and the information storage~\cite{Flassig:2012re}.
These gravitons approximately interact by means of the Newtonian potential
\be
V_{\rm N}(r)
\simeq
-\frac{\gn\,M}{r}
\ ,
\ee
and correspondingly have an effective mass $\mu$ related to their characteristic
quantum-mechanical size via the Compton/de~Broglie wavelength
$\ell \simeq {\hslash}/{\mu} = \lp\,{\mpl}/{\mu}$.
Since they are Bosons, they can superpose in an approximately spherical
volume of radius $\ell$, and total energy $M=N\,\mu$, where $N$ is the total
number of constituent gravitons.
Within the Newtonian approximation~\footnote{One can see that the Newtonian
approximation is not far from the general-relativistic
behaviour~\cite{Ruffini:1969qy,Chavanis:2011cz,Casadio:2013hja}.},
there is then a value of $N$ for which the whole system becomes a black hole.
\par
This can be seen by considering the effective gravitational coupling constant
$\alpha =V_{\rm N}(\ell)/N\simeq {\lp^{2}}/{\ell^{2}} = {\mu^{2}}/{\mpl^{2}}$, and estimating
the average potential energy per graviton as
\be
U
\simeq
\mu\,V_{\rm N}(\ell)
\simeq
-N\,\alpha\,\mu
\ .
\label{eq:Udvali}
\ee
Black hole components in the depleting region are ``marginally bound'' when
\be
E_K + U
\simeq
0
\ ,
\label{eq:energy0}
\ee
where the kinetic energy is given by $E_K\simeq \mu$.
This energy balance yields the ``maximal packing'' condition
\be
N\,\alpha
\simeq
1
\ ,
\label{eq:maxP}
\ee
from which the effective Boson mass $\mu$ and total mass $M$ of the
black hole are seen to scale according to the relations
\be
\mu
\simeq
\frac{\mpl}{\sqrt{N}}
\quad
{\rm and}
\quad
M
=
N\,\mu
\simeq
\sqrt{N}\,\mpl
\ .
\label{eq:Max}
\ee
The horizon's size $\Rh$ and area $A$ are therefore quantized as commonly
expected~\cite{Bekenstein:1997bt},
\be
\Rh
\simeq
\sqrt{N}\, \lp
\quad
\Rightarrow
\quad
A
\sim
\Rh^{2}
\simeq
N\, \lp^{2}
\ .
\label{eq:areaquantization}
\ee
Since this initial intuitive model, other authors have proposed possible
developments~\cite{Casadio:2013hja,Kuhnel:2014oja,mueckPT,Flassig:2012re,Hofmann:2014jya}
and possible cosmological implications~\cite{Binetruy:2012kx, Kuhnel:2014gja}.
\par
The Hawking radiation and the negative specific heat
spontaneously result from the quantum depletion of the condensate
for the states satisfying Eq.~\eqref{eq:energy0}.
At first order, because of reciprocal $2 \to 2$ scattering inside the condensate,
the depletion rate can be estimated to be
\be
\Gamma
\sim
\frac{1}{N^{2}}\,N^{2}\,\frac{1}{\sqrt{N}\,\lp}
\ ,
\label{eq:Grate}
\ee
where the first factor ($N^{-2}$) comes from the interaction strength $\alpha$,
the second factor is combinatoric (there are about $N$ gravitons
scattering with other $N - 1 \simeq N$ gravitons) and the last factor comes from
the characteristic energy of the process.
The amount of gravitons in the condensate will then decrease according
to~\cite{DvaliGomez}
\be
\dot N
\simeq
- \Gamma
\simeq
-\frac{1}{\sqrt{N}\, \lp} + \mathcal{O} (N^{-1})
\label{eq:depleted-N}
\ .
\ee
As explained in Ref.~\cite{DvaliGomez}, such emission of gravitons reproduces
a part of the Hawking radiation which is purely gravitational and contributes to
the shrinking of the black hole.
The usual results are then recovered in the double-scaling limit $N \to \infty$ and $\lp \to 0$,
with $\hbar$ kept finite.
It is important to remark that, in the present description, particle creation by a
black hole is not a vacuum process but an actual scattering phenomenon of the
back hole constituents from the ground state to excited states which display an
``effective'' Hawking temperature.
\par
Despite its simplicity, the lack of a geometrical description of space-time makes
it difficult to understand the role of any kind of horizon in this model.
A possible way to connect the BEC approach to the usual geometrical
point-of-view of general relativity was later proposed~\cite{Casadio:2014vja},
using the formalism of Refs.~\cite{HWF}, which introduces the idea of a ``fuzzy''
gravitational radius at the quantum level, starting from the Einstein equation
that determines the classical Misner-Sharp mass for spherically symmetric systems.
For example, if the quantum state of a single particle is given by a localised
wave-packet obtained from the superposition of many energy eigenstates,
to each eigenstate there will correspond a different gravitational radius
$\Rh$, with a probability amplitude given by the corresponding spectral
coefficient.
When measuring the particle's position, we could then observe it
with some probability inside the expectation value of $\Rh$.
Simple evaluations showed that only very massive particles, of order of the
Planck mass or larger, are likely to appear in a black-hole state, but the relative
uncertainty in the horizon size is then unacceptably large for semiclassical
black holes.
However, when we likewise analyse a system made of many light condensed
Bosons~\cite{Casadio:2014vja}, the horizon relative uncertainty decreases
with $N$, which supports the idea that large BECs of gravitons at the critical point
can be viewed as semiclassical black holes in the large-$N$ limit\footnote{If
one further considers a scalar field $\phi$ coupled to a source $J$ proportional
to the field itself, then Eq.~\eqref{eq:areaquantization} is also
recovered~\cite{Casadio:2014vja}.}.

\section{Quantum Inflationary Cosmology}
\label{sec:Quantum-Inflationary-Cosmology}
\setcounter{equation}{0}

\noindent
As we have already mentioned, the BEC model can be easily extended
to cosmology, and in this Section we systematically review the scenario introduced
in Refs.~\cite{Dvali:2013eja,Dvali:2014gua}.
We shall start by noting that, in order to obtain physically sensible results, one first
needs to reproduce the usual background dynamics, as it is described by the
Friedmann equation
\be
H^{2}
\simeq
\frac{ 8\,\pi }{ 3 }\,\frac{\lp}{\mpl}\,\rho_{\phi}
\ ,
\label{eq:effF}
\ee
at least to some approximation.
We shall in particular show how one can recover separately the
left-hand side (an effective Hubble parameter) and the right-hand side
(the matter source) of the above Friedmann equation.
The former will be effectively given by the graviton condensate, whereas
the latter will require introducing an inflaton field $\phi$, which we shall
further assume satisfies the slow-roll condition.
Next, we shall show that including purely quantum effects 
leads to a master equation for the time-evolution which will be
investigated quantitatively in Sec.~\ref{sec:Cosmological-Parameters}.

\subsection{De Sitter and the Hubble Parameter}
\label{sec:De-Sitter-and-the-Hubble-Parameter}

\noindent
First of all, we note that an observer in the (homogeneous and isotropic)
classical de~Sitter Universe would see a horizon located at the radius
\be
\Rh = H^{-1}
\ ,
\label{eq:RhH}
\ee
where $H$ is the Hubble expansion rate.
Like for black holes~\footnote{Curiously enough, it seems that the radius of the observable
Universe coincides with its Schwarzschild radius (to a good approximation).},
this radius will determine the approximate wavelength and occupation
number of the gravitons in the BEC which make up the space-time fabric.
In fact, let us assume the total gravitational energy confined inside $\Rh$
is again $E=N\,\mu$, where $N$ is the total number of gravitons and
$\mu \sim {\hslash}/{\Rh}\sim \mpl\,\lp/\Rh$ is the single graviton energy.
(Of course, we are here neglecting higher-order corrections to this mean field
approximation, corresponding to terms that scale with lower powers of $N$.)
If we now assume the energy density of this BEC gives rise to an effective
curved geometry with the Hubble parameter~\footnote{For a field-theoretic justification
of this step, see Sec.~3 of Ref.~\cite{Dvali:2013eja}.}
\be
	\frac{ 8\,\pi }{ 3 }\,\frac{\lp}{\mpl}\,\rho
		=
								H^{2}
								\ ,
								\label{eq:friedG}
\ee
one must then have
\be
	H^{2}
		\simeq
								\frac{ 8\,\pi }{ 3 }\,\frac{\lp}{\mpl}\,\frac{E}{\Rh^{3}}
		\simeq
								\frac{ 8\,\pi }{ 3 }\,\frac{\lp}{\mpl}\,\frac{N\,\mu}{\Rh^{3}}
		\sim
								\frac{N}{\Rh^{4}}
								\ .
\ee
From Eq.~\eqref{eq:RhH}, we again obtain
\be
\Rh
\simeq
\sqrt{N}\, \lp
\ ,
\label{eq:Rhc}
\ee
which shows that the critical scalings~\eqref{eq:Max} also hold in the
cosmological case.
The smaller the graviton number in the Hubble patch, the stronger the
expansion ($N \propto H^{-2}$).
Also, the average graviton density
\vs{-2mm}
\be
\rho
\simeq
\frac{M}{\Rh^{3}}
=
\frac{\mpl}{N\, \lp^{3}}
\ ,
\label{eq:rhoN}
\ee
is qualitatively the same as for a black hole, whose mass is quantised
according to Eq.~\eqref{eq:Max}, and where to a larger $N$
corresponds a bigger, less dense and more classical black hole.
\par
Let us remark that another point in common between black holes and
the Friedmann models is that both have some affinity with Newtonian
gravity.
In fact, the full Friedmann equation for the scale factor $a = a( t )$ reads
\be
\dot{a}^{2} - \frac{2\,G_{\Nrm}\, C(a)}{3\,a}
=
k
\ ,
\label{eq:NewtonianHubble}
\ee
where $C \defas 4\,\pi\,\rho\,a^{3}$ is constant for dust, and $k$ is the spatial curvature.
Eq.~\eqref{eq:NewtonianHubble} looks exactly like the equation for the conservation
of the energy $E = 2\,\mu\,k$ of a point particle moving along the trajectory $r = a( t )$
in an external Newtonian potential with source mass $M = C( r )$.
With $k = 0 = E$ for the spatially-flat Universe we seem to live in, this
equation then yields the same marginally bound condition of Eq.~\eqref{eq:energy0}.
It is thus no surprise that the de~Sitter Universe looks like a black hole turned
inside out, where the horizon acts like a semipermeable membrane that prevents
things from re-entering, whereas in a black hole it keeps them from escaping.
\par
As shown by Dvali and Gomez, there is one important difference between black
holes and cosmology that we need to take into account now:
the passage from de~Sitter to Minkowski forces us to
\emph{``introduce a cosmological fluid that plays the role of a homogeneous
cosmic clock.
This role is played by a scalar field, the inflaton $\phi$''}~\cite{Dvali:2013eja}.
In other words, having recovered the left-hand side of Eq.~\eqref{eq:effF},
we now need to have a suitable matter source in its right-hand side.

\subsection{Slow-Roll Inflation}
\label{sec:Slow--Roll-Inflation}

\noindent
In light of the above observation, let us go back to old-fashioned general relativity
and consider a homogeneous inflaton scalar field $\phi$ of mass $m = \hslash\,\lambda_{m}^{-1}$
and with the simple potential
\vs{-2mm}
\begin{align}
	V( \phi )
		&=
								\frac{ 1 }{ 2 }\.\lambda_{m}^{-2}\,\phi^{2}
								\; .
								\label{eq:V}
\end{align}
The energy density and the pressure are given by
\be
	\rho_{\phi}
		=
								\frac{1}{2}
								\left(
									\dot \phi^{2}
									+
									\lambda_{m}^{-2}\,\phi^{2}
								\right)
		\qq
		\text{and}
		\qq
	p_{\phi}
		=
								\frac{1}{2}
								\left(
									\dot \phi^{2}
									-
									\lambda_{m}^{-2}\,\phi^{2}
								\right)
								,
								\label{eq:Friedmann-Equation}
\ee
respectively.
The field $\phi$ evolves according to the Klein-Gordon equation
\be
	\ddot \phi
	+
	3\, H\, \dot \phi
	+
	\lambda_{m}^{-2}\, \phi
		=
								0
								\ .
\ee
In the slow-roll approximation, $|\overset{.}{\phi}{}^{2}| \ll | V |$ and
$|\overset{..}{\phi}{}^{2}| \ll 3\.H\.\overset{.}{\phi} \sim | V_{,\phi} |$, the inflaton therefore
displays a quasi-de~Sitter-vacuum equation of state
\be
p_{\phi}
\simeq
-\rho_{\phi}
\simeq
-\frac{1}{2}\,\lambda_{m}^{-2}\,\phi^{2}
\ .
\ee
This leads to the de~Sitter exponential growth of the Universe, with
\be
\frac{ 4\,\pi }{ 3\.\lambda_{m}^{2} }\frac{\lp}{\mpl}\,\phi^{2}
\simeq
H^{2}
\ .
\ee
The slow-roll parameter
\be
	\epsilon
		\defas
								-
								\frac{\dot H}{H^{2}}
		\simeq
								\frac{1}{3 \left(\lambda_{m}\,H\right)^{2}}
		=
								\frac{ 1 }{ 3 }\!
								\left(
									\frac{ R_{\rm H} }{ \lambda_{m} }
								\right)^{2}
		\ll
								1
\ ,
\ee
determines the validity of the approximation, and shows that the Compton length
of the inflaton must be much larger than the Hubble radius.
In particular, we can take the common assumption that the exponential inflation
turns off when $\epsilon$ approaches 1, or $\lambda_{m}\simeq \Rh$.
\par
In the slow-roll approximation, the total inflaton occupation
number inside the Hubble patch can be estimated by multiplying the inflaton
number density,
\be
	n_{\phi}
		=
								\frac{\rho_{\phi}}{m}
		\simeq
								\frac{ 3 }{ 8\,\pi }\,\frac{\lambda_{m}}{\lp^{2}}\,H^{2}
								\ ,
								\label{eq:nphi}
\ee
times the Hubble volume $\Rh^{3} \simeq 1/H^{3}$.
By simply substituting in for the previous quantities, and recalling Eq.~\eqref{eq:RhH},
we find the total number of inflatons is given by
\be
	N_{\phi}
		\simeq
								n_{\phi}\,\Rh^{3}
		\simeq
								\frac{ 3 }{ 8\,\pi }\,\frac{\lambda_{m}}{\lp^{2}}\,H\,\Rh^{2}
								\ .
								\label{eq:Nphi}
\ee
Using Eq.~\eqref{eq:Rhc}, one can then see that the total inflaton number is related
to the total graviton number via the Hubble constant, that is
\be
	N_{\phi}
		\simeq
								\left(
									\frac{ 3 }{ 8\,\pi }\,\lambda_{m}\,H
								\right)
								N
		\simeq
								\frac{ \sqrt{3\,} }{ 8\,\pi }\;\frac{N}{\sqrt{\epsilon}}
								\ ,
								\label{eq:NphiN}
\ee
therefore one finds that $N_{\phi} \gg N$ in the slow-roll regime.
\par
Let us now abandon the standard general-relativistic description of cosmology,
and return to the BEC model of Sec.~\ref{sec:De-Sitter-and-the-Hubble-Parameter}.
We then see that the above relation~\eqref{eq:NphiN} between $N$ and $N_{\phi}$
can be viewed as the effective background Friedmann equation~\eqref{eq:effF},
that is
\be
\frac{ 8\,\pi }{ 3 }\,\frac{\lp}{\mpl}\,\rho
\simeq
H^{2}
=
\frac{ 8\,\pi }{ 3 }\,\frac{\lp}{\mpl}\,\rho_{\phi}
\ ,
\label{eq:Fried}
\ee
where $\rho\simeq \mpl/N\,\lp^{3}$ is the energy density~\eqref{eq:rhoN} of the
gravitons in the BEC and $\rho_{\phi}\simeq N_{\phi}\,m\,H^{3}$ is precisely the inflation
energy density in Eq.~\eqref{eq:nphi}.
Although one must have $N\ll N_{\phi}$ in the slow-roll regime, the energy densities
of gravitons and inflatons remain equal, which reminds us that the above effective
Friedmann equation is just a statement of total energy conservation~\footnote{In
general relativity, Eq.~\eqref{eq:effF} is the Hamiltonian constraint that
follows from the Einstein-Hilbert action.}.
\par
To summarise, we have so far shown how to recover purely classical background dynamics
compatible with the Friedmann equation~\eqref{eq:effF}.
It is then time to include quantum effects, namely the depletion of the
background BEC that, for black holes, leads to the Hawking evaporation.

\subsection{Quantum Time Evolution}
\label{sec:Quantum-Time-Evolution}

\noindent
Like inside a black hole, we expect that scattering of gravitons will lead to a progressive
depletion of the background BEC, now further enhanced by the scattering against inflatons
(see Ref.~\cite{Dvali:2013eja} for more details).
If we further assume Eq.~\eqref{eq:Fried} still holds (to leading order in $N$),
given the depletion law for the $N$ gravitons, we obtain the time-evolution of
$N_{\phi}$, which is like implicitly defining $\dot \phi$, and therefore finding a way
to turn the inflation off.
One could also introduce a wave-function for the Hubble radius by quantising the
Friedmann equation~\eqref{eq:Fried}, like one does with the Misner-Sharp mass of
a compact source to describe the Schwarzschild radius of quantum black holes.
The same formalism would yield a ``fuzzy'' cosmological horizon (and therefore $H$), with an uncertainty
that we can anticipate will decrease for large $N$ (and, correspondingly, large $N_{\phi}$).
In fact, we shall apply this formalism just to show the effect of excited modes
on the background evolution remains negligibly small within the slow-roll regime
(see Appendix~\ref{app:Effect-of-Non--Condensed-Quanta}).
More importantly, we remark that, like the depleted quanta make up the Hawking
radiation from black holes, the excited inflatons will appear to an observer as the
scalar density perturbations, and the excited gravitons as the tensor perturbations.
Like for black holes, in this model there is no production of quanta from the vacuum, and
the fact that the higher energy of the quanta above the ground state does not significantly
affect the background dynamics can be translated into the statement that the backreaction
of density and tensor perturbations is mostly negligible.
\par
In order to study the time evolution of our inflationary background, we can directly
employ the relation~\eqref{eq:friedG} between the Hubble rate $H$ and
the energy density $\rho$ of the gravitons.
The novelty with respect to Section~\ref{sec:Slow--Roll-Inflation}, where the inflaton was introduced
in order to recover the slow-roll evolution, is that we also want to take into proper account
the effect due to the depletion of the BEC.
In particular, the master equation for the number $N$ of gravitons in the BEC used in
Ref.~\cite{Dvali:2013eja} can be obtained rather straightforwardly.
\par
The rate of change of $N$ will in fact contain two contributions:
a ``classical'' (mean field) one due to the presence of the inflaton and described
by the (effective) Friedmann equation, and a purely ``quantum'' one due to the
depletion of the background BEC,
\be
\frac{\dot N}{N}
=
\frac{\dot N_{\rm cl}}{N} + \frac{\dot N_{\rm dep}}{N}
\ .
\ee
The former contribution is obtained via the Friedmann equation~\eqref{eq:Fried}
and Eq.~\eqref{eq:rhoN}, that is
\be
-\frac{\dot N_{\rm cl}}{N}
\simeq
\frac{\dot\rho}{\rho}
=
\frac{1}{H^{2}}\,\frac{\d H^{2}}{\d t}
=
2\,\frac{\dot H}{H}
=
-2\, \epsilon\, H
\ .
\ee
The quantum depletion is again given by $\dot N_{\rm dep} \sim -\Gamma$,
where $\Gamma$ is obtained from the expression in Eq.~\eqref{eq:depleted-N} with
a small change to account for the presence of more species.
As explained in Ref.~\cite{Dvali:2013eja}, we must consider three kinds of interactions:
graviton-graviton ($g-g$), graviton-inflaton ($g-\phi$) and inflaton-inflaton ($\phi-\phi$).
Since the inflaton wavelength is virtually infinite in the slow-roll approximation
($\lambda_{m} \gg R_{\rm H}$),
the $\phi-\phi$ process is irrelevant with respect to the scatterings involving gravitons.
The $g-g$ and $g-\phi$ scatterings differ by a combinatoric factor $N_{\phi}$,
which is by far larger than $N$ in the slow-roll regime [see Eq.~\eqref{eq:NphiN}].
It is therefore the $g-\phi$ process that dominates and depletes gravitons and
inflatons at the same rate $\Gamma_{g\phi}$.
Taking furthermore into account that the characteristic momentum transfer is
$\hslash\,H\simeq \lp\,\mpl\,H$, we find that the depletion rate $\Gamma_{g\phi}$
can be well approximated by (\cf~Eq.~(5.12) in Ref.~\cite{Dvali:2013eja})
\begin{align}
	\Gamma_{g\phi}
		&\simeq
								\alpha^{2}\.N_{\phi}\.N\.H
								\; .
								\label{eq:Gamma-g-phi}
\end{align}
Here, the first factor (the square of the dimensionless self-coupling
$\alpha = (\lp\, H)^{2}=(\lp/\Rh)^{2}$) stems from the fact that the dominant
process is again give by $2\to 2$ scatterings.
The factor $N_{\phi}\,N$ counts the available pairs in the $g-\phi$ scattering,
and $H$ is the characteristic momentum transfer, as mentioned above.
Using the ``maximal packing'' condition $\alpha\.N \simeq 1$, we have
\be
	\dot N
		=
								\dot N_{\phi}
	\simeq
								-
								\Gamma_{g\phi}
	\simeq
								-
								H \,\alpha^{2}\, N_{\phi}\, N
	=
								-
								\frac{1}{\sqrt{N}\,\lp}\,\frac{N_{\phi}}{N}
								\label{eq:gammagf}
								\ .
\ee
With Eq.~\eqref{eq:NphiN}, this leads to
\be
	\frac{\dot N_{\rm dep}}{N}
		\simeq
								-
								\frac{\Gamma_{g\phi}}{N}
		\simeq
								-
								\frac{1}{\sqrt{N}\,\lp}\,\frac{N_{\phi}}{N}\, \frac{1}{N}
		=
								-
								\frac{ \sqrt{3\,} }{ 8\,\pi }\.\frac{H}{\sqrt{\epsilon}N}
								\ .
								\label{eq:minus-Ndot-classical-over-N}
\ee
Putting together these pieces, and temporarily omitting numerical factors
in order to highlight the functional dependence on $N$,
we finally find the master equation (\cf~Eq.~(5.27) in Ref.~\cite{Dvali:2013eja})
\vs{-2mm}
\be
	\frac{\dot N}{N}
		\simeq
								\epsilon\, H
								-
								\frac{H}{\sqrt{\epsilon}\,N}
								\ .
								\label{eq:master-equation}
\ee
After replacing $\epsilon \simeq m^{2} / ( 3\. H^{2}\. \lp^{2}\. \mpl^{2} )$
and $H = \Rh^{-1} \simeq (\sqrt{N}\,\lp)^{-1}$,
and making the time-dependence manifest, we get
\be
	\dot{N}( t )
		=
								\frac{ 2 }{ 3 }\,\frac{m^{2}\,N^{3 / 2}( t )}{\mpl^{2}\, \lp}
								-
								\frac{ \mpl }{ \lp\,m\,N( t ) }
								\ ,
								\label{eq:N-dot}
\ee
which reproduces the classical Friedmann dynamics in the double-scaling limit $N \to \infty$
and $\hbar \to 0$, with $G_{\Nrm}$ and $m$ kept finite.

Corrections arise when going to higher orders in $1 / N$, which find correspondence in higher orders in the BBGKY-hierachy \cite{BBGKY}. The influence of the first of such terms, \ie~the depletion term in Eqs.~\eqref{eq:master-equation} and \eqref{eq:N-dot}, gives rise to quantum corrections, which differ from the standard semi-classical results to which we will compare our respective findings below. All these corrections cumulate, and higher-order terms will have to be taken into account after a certain time. This is precisely what limits the total duration of inflation (\cf~Sec.~5.4 of Ref.~\cite{Dvali:2013eja}) as the mean-field (background) description becomes poorer and poorer. Then, eventually, one would need to go beyond the first $1 / N$ term, or, respectively, the second-order truncation of the BBGKY-hierachy.

We again remark that we have simply retained the functional dependence on $m$ and $N$, and set a numerical
coefficient in front of the second term on right-hand side equal of Eq.~\eqref{eq:N-dot} to one.
This simplification should be kept in mind when discussing the results in the
next Section, since a different relative numerical coefficient between the two
contributions to the evolution of $N$ will likely change the final figures, but this
analysis is left for future investigations.
%
%
%
\section{Cosmological Parameters}
\label{sec:Cosmological-Parameters}
\setcounter{equation}{0}
\noindent
Of course, Eq.~\eqref{eq:N-dot} has to be solved with appropriate boundary conditions,
which we will choose to specify at the end of inflation $t = t_{\rm end}$,
i.e.~$N( t_{\rm end} ) = N_{\star}$, for a certain number of condensed background gravitons $N_{\star}$.
This will be estimated, given the energy scale $m$, by $N_{\star} \sim ( \mpl / m )^{2}$;
for instance for $m = 10^{13}\,{\rm GeV}$ one has $N_{\star} \sim 10^{12}$.
We then solve Eq.~\eqref{eq:N-dot} numerically and backwards from $t_{\rm end}$
over a time interval $[ t_{\rm start},\.t_{\rm end} ]$ corresponding to a total number
$N_{\erm} = 200$ of e-foldings.
For the simple inflaton potential \eqref{eq:V} one has
\be
N_{\erm} = \frac{m^{2}}{6}\.( t_{\rm start} - t_{\rm end} )^{2}
\ ,
\ee
where the time $t_{\rm end}$ is characterised by the condition
$\epsilon( t_{\rm end} ) = 1$.
\par
\begin{figure}
	\centering
	\includegraphics[scale=1.40,angle=0]{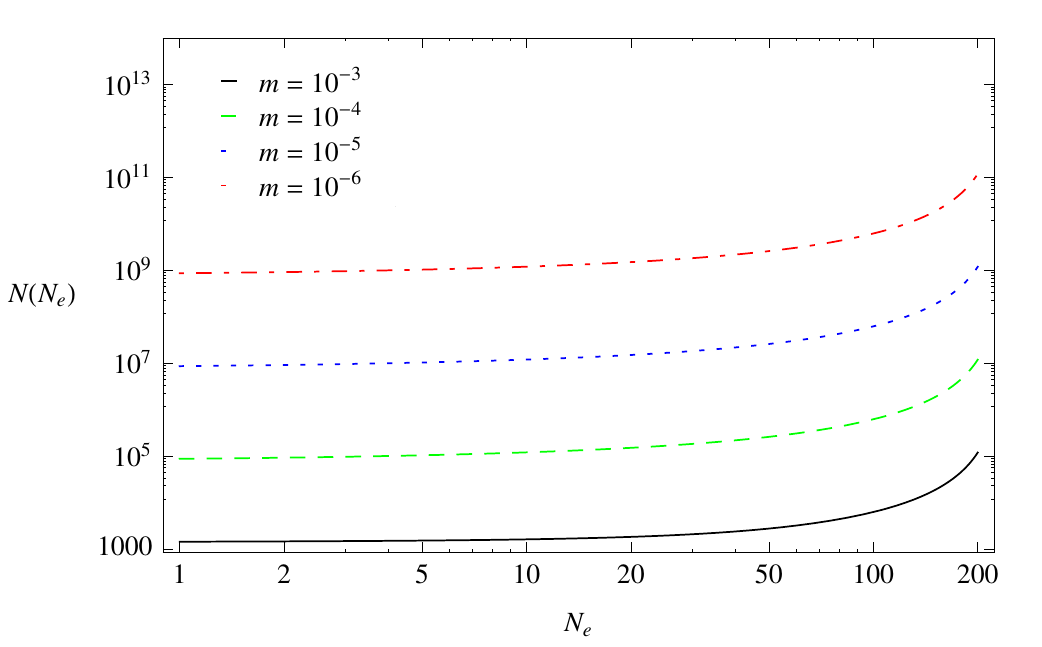}
	\includegraphics[scale=1.40,angle=0]{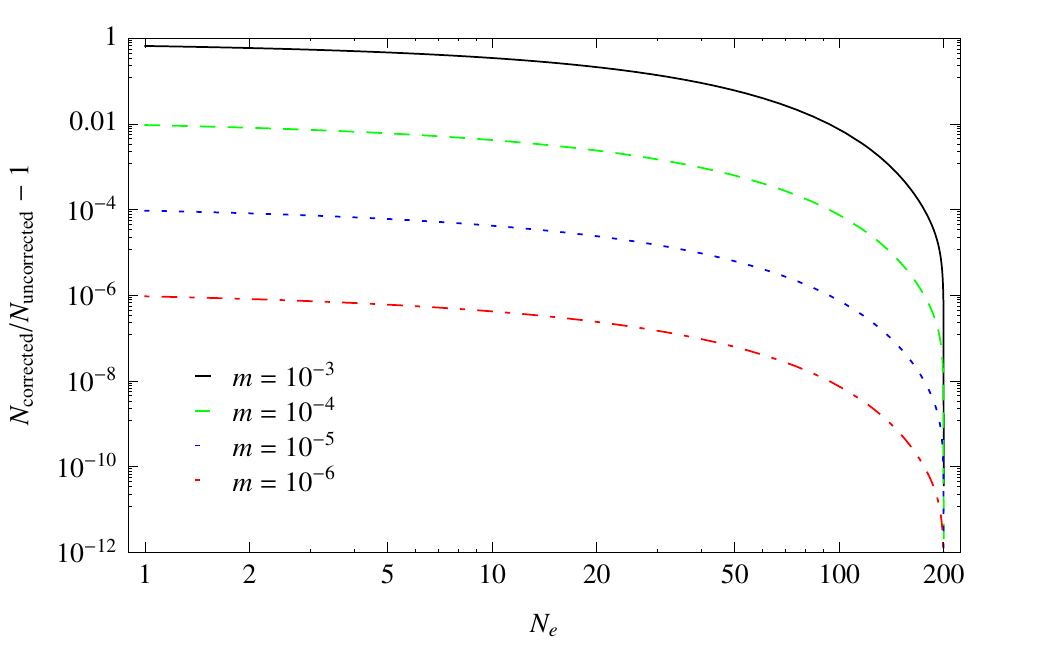}
	\caption{{\it Upper panel:}
			The number of condensed gravitons $N( N_{\erm} )$ as a function of the number of e-foldings
			for various values of the inflaton mass $m = 10^{-3,\,\ldots,\,-6}\,\mpl$ (bottom to top).
			{\it Lower panel:}
			The relative correction $N_{\rm corrected}( N_{\erm} ) / N_{\rm uncorrected}( N_{\erm} ) - 1$
			of the quantum-corrected to classical evolution of the number of condensed
			gravitons as a function of the number of e-foldings,
			for various values of the inflaton mass $m = 10^{-3,\,\ldots,\,-6}\,\mpl$ (top to bottom).
			Masses are in units of $\mpl$.}
	\label{fig:N}
\end{figure}

The upper panel of Fig.~\ref{fig:N} shows the time-dependence of the number
of condensed gravitons for various values of the inflaton mass, and
we display the relative correction of the full solution to Eq.~\eqref{eq:N-dot}
with respect to the one without quantum depletion in the bottom panel
of the same Fig.~\ref{fig:N}.
We observe that the shapes for the different values of the mass
are very similar, with the quantum-depletion correction becoming increasingly
important for increasing mass.
In fact, this correction roughly scales as $m^{2}$.
\footnote{Of course, for large-enough masses and for a sufficiently long
duration of inflation, the underlying approximations will always break down at
some point, as quantum-depletion effects continuously and cumulatively lower
the amount of condensed quanta, making successively the $1/N$-corrections
more and more important.
The larger the mass the sooner this will happen.
However, for all the cases studied in this work, those corrections are under control.}
\par
\begin{figure}
	\centering
	\includegraphics[scale=1.40,angle=0]{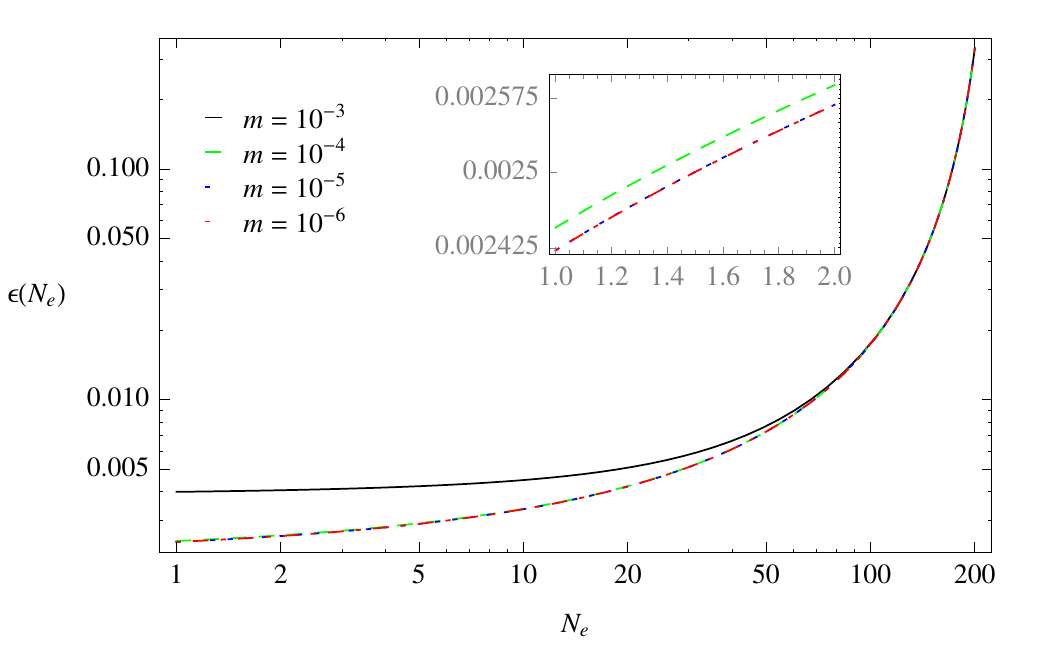}
	\includegraphics[scale=1.40,angle=0]{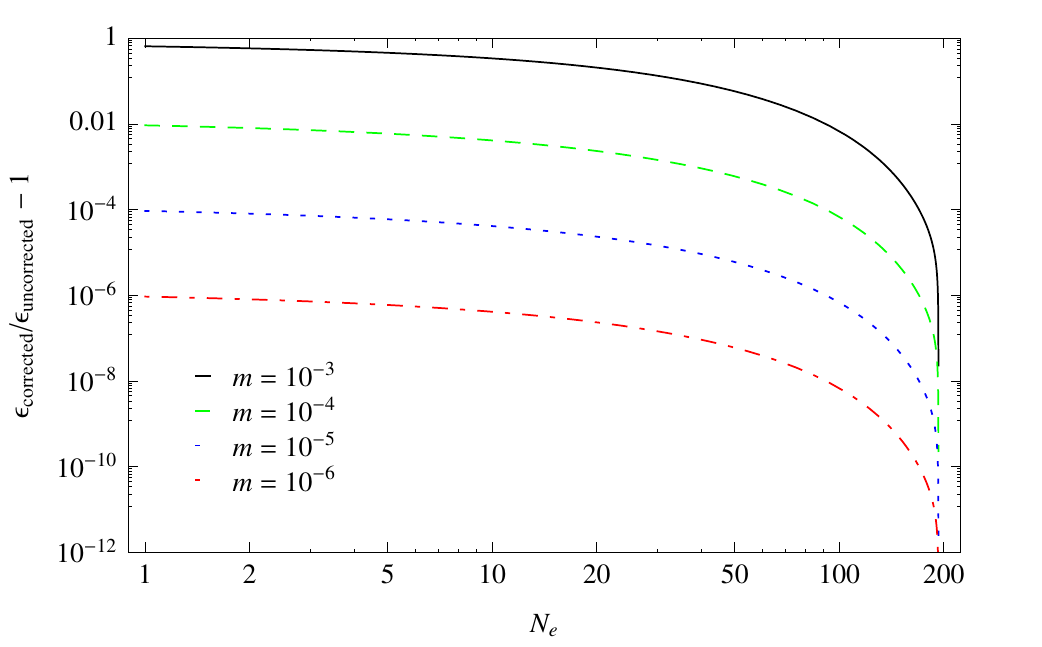}
	\caption{{\it Upper panel:}
			The slow-roll parameter $\epsilon( N_{\erm} )$ as a function of the number of e-foldings
			for various values of the inflaton mass $m = 10^{-3,\,\ldots,\,-6}\,\mpl$ (top to bottom).
			{\it Lower panel:}
			The relative correction $\epsilon_{\rm corrected} / \epsilon_{\rm uncorrected} - 1$
			of the quantum corrected to the classical evolution of the slow-roll parameter
			$\epsilon( N_{\erm} )$ as a function of the number of e-foldings,
			for various values of the inflaton mass $m = 10^{-3,\,\ldots,\,-6}\,\mpl$ (top to bottom).
			Masses are in units of $\mpl$.}
	\label{fig:Epsilon}
\end{figure}
\begin{figure}
	\centering
	\includegraphics[scale=1.40,angle=0]{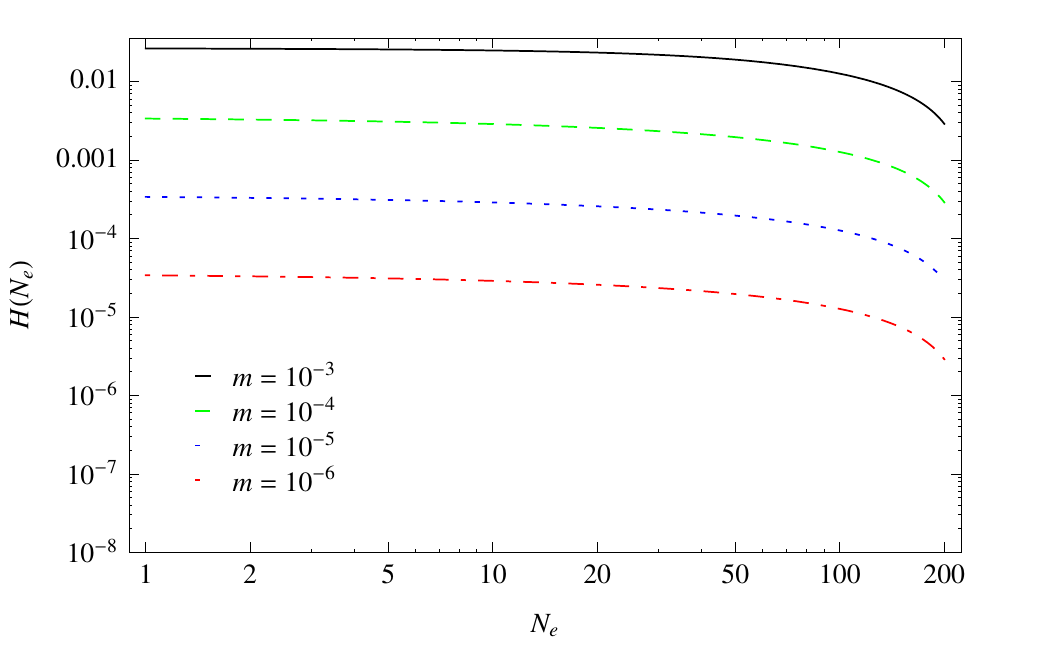}
	\includegraphics[scale=1.40,angle=0]{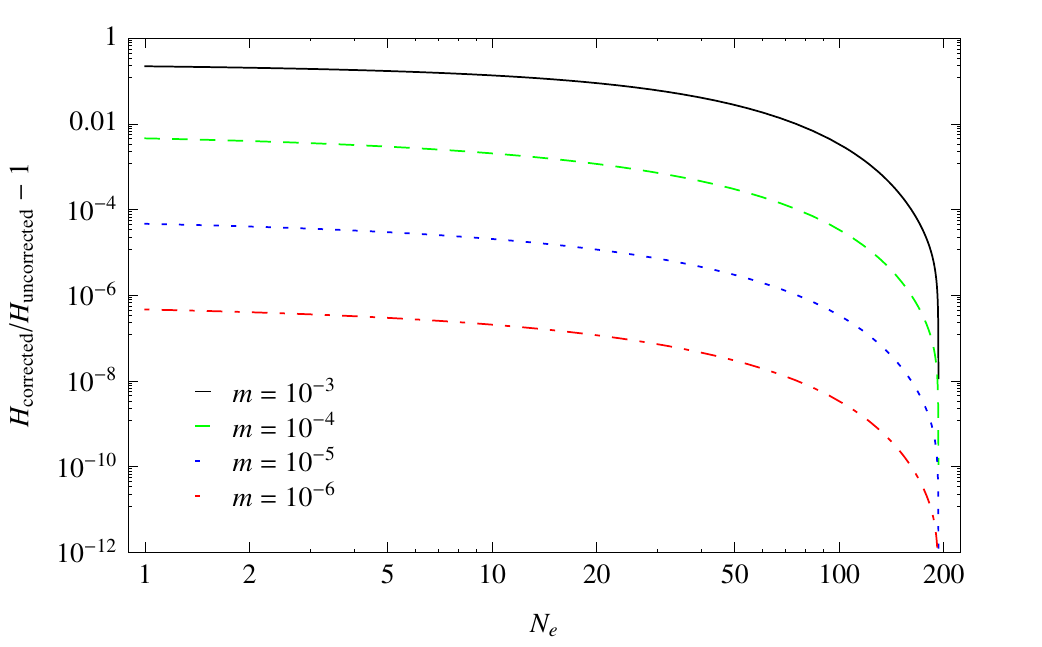}
	\caption{{\it Upper panel:}
			The Hubble parameter $H( N_{\erm} )$ as a function of the number of e-foldings
			for various values of the inflaton mass $m = 10^{-3,\,\ldots,\,-6}\,\mpl$ (top to bottom).
			{\it Lower panel:}
			The relative correction $H_{\rm corrected} / H_{\rm uncorrected} - 1$
			as a function of the number of e-foldings,
			for various values of the inflaton mass $m = 10^{-3,\,\ldots,\,-6}\,\mpl$ (top to bottom).
			Masses are in units of $\mpl$.}
	\label{fig:HRatio}
\end{figure}

The top panel of Fig.~\ref{fig:Epsilon} displays the slow-roll parameter $\epsilon( N_{\erm} )$
as a function of the number of e-foldings $N_{\erm}$.
It is easy to see that $\epsilon$ is larger for larger mass $m$ at earlier times,
and it then approaches one at the end of inflation
(which corresponds to $N_{\erm} = 200$ here).
Similarly to Fig.~\ref{fig:N}, we plot in the bottom panel of Fig.~\ref{fig:Epsilon}
the corresponding relative correction $\epsilon_{\rm corrected} / \epsilon_{\rm uncorrected} - 1$,
which is also approximately proportional to $m^{2}$.
The same holds true for the relative correction $H_{\rm corrected} / H_{\rm uncorrected} - 1$
of the Hubble parameter (which scales like $1 / \sqrt{N}$), being presented in Fig.~\ref{fig:HRatio}.

At this point, one might wonder about the corrections to the tensor-to-scalar ratio,
since it is strongly sensitive to the inflaton mass, but is also expected to undergo
quantum-depletion corrections.
This is even more true as it probes the {\it total\/} duration of the inflationary expansion,
and is approximately given by (\cf~Eq.~(5.36) of Ref.~\cite{Dvali:2013eja})
\begin{align}
	r( t )
		&\simeq
								\Bigg(
									\frac{ N( t ) }{ N_{\phi}( t ) }
								\Bigg)^{\!\!2}\.
								\Bigg[
									1
									+
									\frac{ 1 }{ N( t ) }
									\int_{t_{\rm start}}^{t_{\rm end}}\d \tau\; \frac{ H( \tau ) }{ \sqrt{\epsilon( \tau )\,} }
								\Bigg]
								\; .
								\label{eq:r}
\end{align}
For evaluating this quantity we employ results from the Planck paper~\cite{Ade:2013uln}; 
we take the pivot scale $k_{\rm pivot} = 0.05\.{\rm Mpc}^{-1}$,
and compute $r( t_{\rm pivot} )$ at the time $t_{\rm pivot}$ when $k_{\rm pivot}$
exited the horizon.
Of course, as the Hubble parameter is affected by quantum depletion, we
remark that $t_{\rm pivot}$ also acquires a correction.

For the potential used in this work, the Planck collaboration gives the upper bound
$r_{0.05} \lesssim 0.12$
(Planck+WP $95\.\%$ CL, \cf~Tab.~9 in Ref.~\cite{Ade:2013uln})~\footnote{The bound
obtained by including the running of the spectral index is $r < 0.25$
($\Lambda{\rm CDM} + r + \d n_{s} / \d \ln( k )$, Planck+WP, \cf~Tab.~5 in Ref.~\cite{Ade:2013uln})}.
More recently, this bound was confirmed after including BICEP2 and Keck Array
data in Ref.~\cite{Planck15}.
We tabulate the specific values of $r$ at the chosen pivot scale in Tab.~\ref{tab:r}.
One can observe that masses up to $m \approx 10^{-4}\,\mpl$ are consistent with this Planck bound.
Furthermore, we see that, for large masses, the influence of the quantum depletion is rather strong
and it actually {\it lowers\/} the value of $r$, hence, it generically broadens the range of compatible
masses.
This effect is stronger the higher the inflaton mass, as expected since the effect
of quantum depletion increases with increasing mass (\cf~the above discussion).

\if\case
\begin{table}
\begin{center}
\begin{tabular}{|l||c|c|c|c|c|c|c|c|}
\hline
\backslashbox{\;\text{correction}}{\;$m [\mpl]$}
					&\;$10^{-6}$
						$\vphantom{1^{^{^{^{^{^{^{^{^{1}}}}}}}}}}$\;				
					&\;$10^{-5}$\;
					&\;$10^{-4}$\;
					&\;$10^{-3}$\;
					&\;$5 \cdot 10^{-3}$\;
					&\;$10^{-2}$\;
					&\;$1.5 \cdot 10^{-2}$\;
					&\;$2 \times 10^{-2}$\;\\
\hline\hline
	\;\text{uncorrected}
					&\;$0.12$
					&\;$0.12$
					&\;$0.12$
					&\;${\color{red}0.15}$
					&\;${\color{red}0.84}$
					&\;${\color{red}2.98}$
					&\;${\color{red}6.56}$
					&\;${\color{red}11.58}$\\
\hline
	\;\text{quantum depletion}
					&\;$0.12$
					&\;$0.12$
					&\;$0.12$
					&\;${\color{red}0.14}$
					&\;${\color{red}0.36}$
					&\;${\color{red}0.74}$
					&\;${\color{red}1.17}$
					&\;${\color{red}1.65}$\\
\hline
\end{tabular}
\end{center}
\caption{The tensor-to-scalar ratio $r$ at the pivot scale $k_{\rm pivot} = 0.05\,{\rm Mpc}^{-1}$;
		red numbers indicate violations of the bound $r < 0.12$ (see main text).}
\label{tab:r}
\end{table}
\fi

\begin{table}
\begin{center}
\begin{tabular}{|l||c|c|c|c|c|c|c|c|}
\hline
\backslashbox{\;\text{ratio}}{\;$m [\mpl]$}
					&\;$10^{-6}$
						$\vphantom{1^{^{^{^{^{^{^{^{^{1}}}}}}}}}}$\;				
					&\;$10^{-5}$\;
					&\;$10^{-4}$\;
					&\;$10^{-3}$\;
					&\;$5 \cdot 10^{-3}$\;
					&\;$10^{-2}$\;
					&\;$1.5 \cdot 10^{-2}$\;
					&\;$2 \times 10^{-2}$\;\\
\hline\hline
	\;\text{corrected / uncorrected}
					&\;$1$
					&\;$1$
					&\;$1$
					&\;$0.93$
					&\;$0.43$
					&\;$0.25$
					&\;$0.18$
					&\;$0.14$\\
\hline
\end{tabular}
\end{center}
\caption{The tensor-to-scalar ratio $r$ at the pivot scale $k_{\rm pivot} = 0.05\,{\rm Mpc}^{-1}$;
		red numbers indicate violations of the bound $r < 0.12$ (see main text).}
\label{tab:r}
\end{table}

Let us next come back to the quantum-depletion effect on the Hubble parameter.
As can be observed in Fig.~\ref{fig:HRatio}, $H$ will be suppressed more strongly at earlier times.
This immediately leads us to investigate whether we can learn something about the observed
suppression of the low multipoles~\cite{lowmultipole} of the CMB temperature
auto-correlation function.

In general, the spectrum of (relative) temperature fluctuations $\delta{T} / {T}$ can be
described by an infinite set of correlation functions
\begin{align}
	\Ccal\!\left( \theta_{1\.2}, \theta_{1\.3}, \ldots, \theta_{n-1\.n} \right)
		&\defas					\left\langle\frac{\delta{T}}{ {T} }( \lbm_{1} )\,\frac{ \delta{T} }{ {T} }( \lbm_{2} )
								\cdots \frac{ \delta{T} }{ {T} }( \lbm_{n} )
								\right\rangle_{\!\!\theta_{1\mspace{2mu}2},\mspace{2mu}
								\theta_{1\mspace{2mu}3}, \ldots,\mspace{2mu}
								\theta_{n-1\mspace{2mu}n}}
								\label{eq:C-of-multi-theta--definition}
\end{align}
with $n \in \Nbbm$ and the angle brackets denote an average over all possible
directions $\lbm_{i}$, for a given configuration of angles
$\theta_{i\.k} \defas \arccos( \lbm_{i}\cdot\lbm_{k} )$ (see Ref.~\cite{Hu:2001bc} for
an extended review).
Assuming approximate Gaussianity, the spectrum can be described by the two-point function
\begin{align}
	\Ccal\!\left( \theta \right)
		&\defas					\Ccal\!\left( \theta_{1\.2} \right)
		=						\left\langle\frac{\delta{T}}{ {T} }( \lbm_{1} )\,
								\frac{ \delta{T} }{ {T} }( \lbm_{2} )
								\right\rangle_{\!\!\theta_{1\mspace{2mu}2}}
								\label{eq:C-of-theta--definition}
\end{align}
alone.
It is convenient to decompose it as
\begin{align}
	\Ccal\!\left( \theta \right)
		&\equiv					\sum_{\ell = 2}^{\infty} \frac{ 2\,\ell + 1 }
								{ 4 \pi }\,\Crm_{\ell}\,\Prm_{\ell}\!
								\left[ \cos( \theta ) \right]
								,
								\label{eq:C-of-theta--multipole-decomposition}
\end{align}
where the multipole moments $\Crm_{\ell} \in \Rbbm$ are expansion coefficients,
$\Prm_{\ell}$ is a Legendre polynomial of degree $\ell$,
and we subtracted the monopole and the dipole.
Again, under the assumption of homogeneity and isotropy, the average over
directions{\;---\;}appearing in the definition of $\Ccal\!\left( \theta \right)${\;---\;}is
equivalent to the average over different points, keeping the directions
$\lbm_{1}$ and $\lbm_{2}$ fixed with the condition $\theta = \arccos( \lbm_{1}\cdot\lbm_{2} )$,
for a given angle $\theta$.

\begin{figure}
	\centering
	\includegraphics[scale=1.40,angle=0]{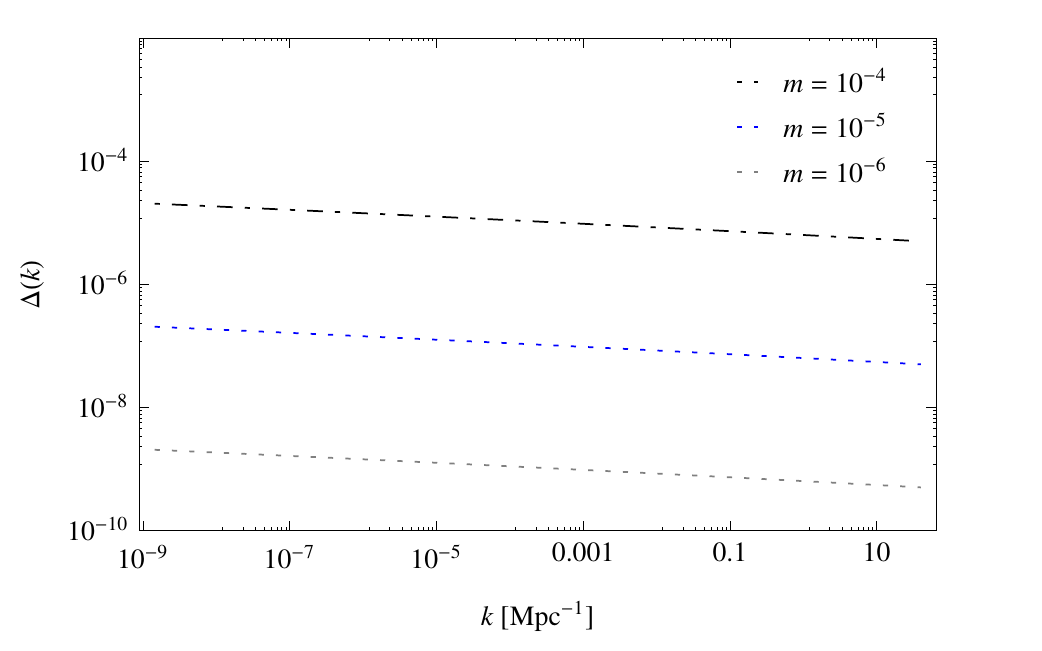}
	\caption{The quantity $\Delta( k ) \defas 1 - H_{\rm corrected}^{2}( k ) / H_{\rm uncorrected}^{2}( k )$
			as a function of the wave-number $k$ for various values of the inflaton mass 
			$m = 10^{-4,\,-5,\,-6}\,\mpl$ (top to bottom).
			Masses are in units of $\mpl$.}
	\label{fig:Delta}
\end{figure}

The multipole moments $\Crm_{\ell}$ are, modulo transfer functions and prefactors,
given by a $k$-integral over the inflationary power spectrum which is approximately
proportional to $H^{2}$ (\cf, \eg, Ref.~\cite{Baumann:2009ds}).
To this end, in Fig.~\ref{fig:Delta}, we plot the quantity 
\be
	\Delta( k )
		\defas
								1
								-
								\frac{H_{\rm corrected}^{2}( k )}{H_{\rm uncorrected}^{2}( k )}
								\, ,
\ee
which shows us exactly how the correction to the integrand of the mentioned $\Crm_{\ell}$-integral behaves.
For small scales it is approximately proportional to $m^{2}$.

\begin{figure}
	\centering
	\includegraphics[scale=1.40,angle=0]{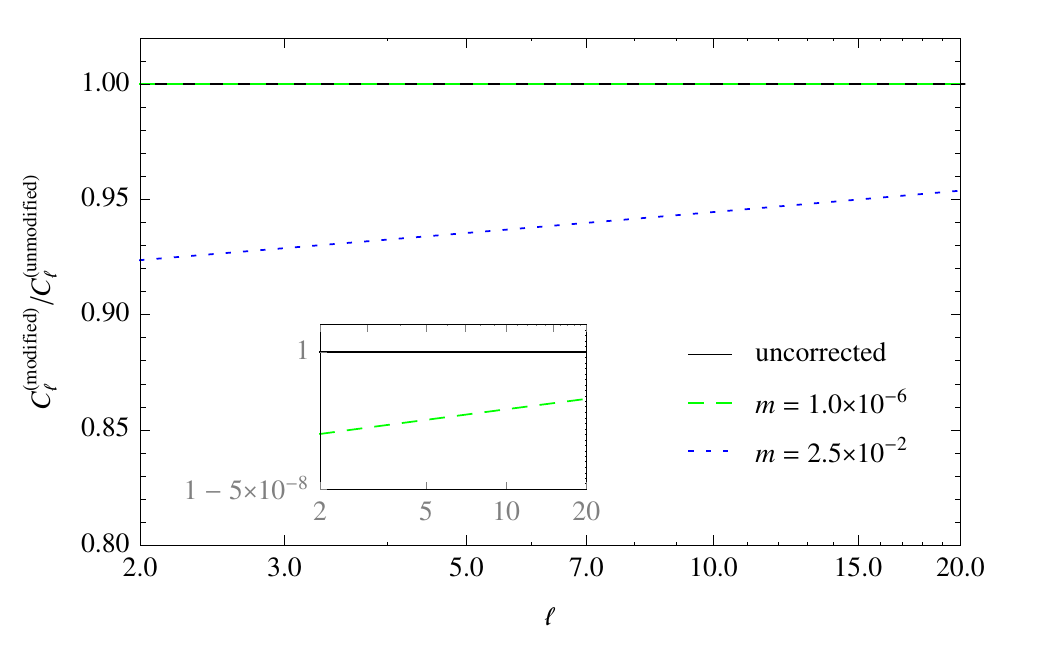}
	\caption{The ratio $\Crm_{\ell}^{\rm corrected} / \Crm_{\ell}^{\rm uncorrected}$ of the uncorrected
			to the quantum-depletion modified multipole moments.
			Masses are in units of $\mpl$.}
	\label{fig:ClRatio}
\end{figure}

For the calculation of the coefficients $\Crm_{\ell}$ we use a modified
version of \texttt{CAMB}~\cite{CAMB}, employing the \texttt{CMBQuick}
package~\cite{CMBQuick}.
Fig.~\ref{fig:ClRatio} shows the ratio of the uncorrected to the quantum-depletion-modified
multipole moments.
The important result here is given by the green dashed line corresponding
to $m = 10^{6}\,\mpl$, which, for the potential~\eqref{eq:V}, is the standard required
value leading to the normalisation of the matter power-spectrum used by the
Planck collaboration~\cite{Ade:2013uln} (\cf~p.~64 in Ref.~\cite{Baumann:2009ds}).
For illustrative purposes only, we also display the relative corrections arising
from the choice of $m = 2.5\.\cdot\.10^{-2}\,\mpl$, which we add to the standard
(properly normalised) result.
Although the base curve onto which the corrections are added would be different,
the resulting line illustrates the behaviour of the respective quantum-depletion correction.
We indeed observe a potentially significant suppression of the multipole moments for small $\ell$,
which corresponds to large angular scales.
As expected, this suppression is stronger for larger values of the inflaton mass.
\par
\begin{figure}
	\centering
	\includegraphics[scale=1.40,angle=0]{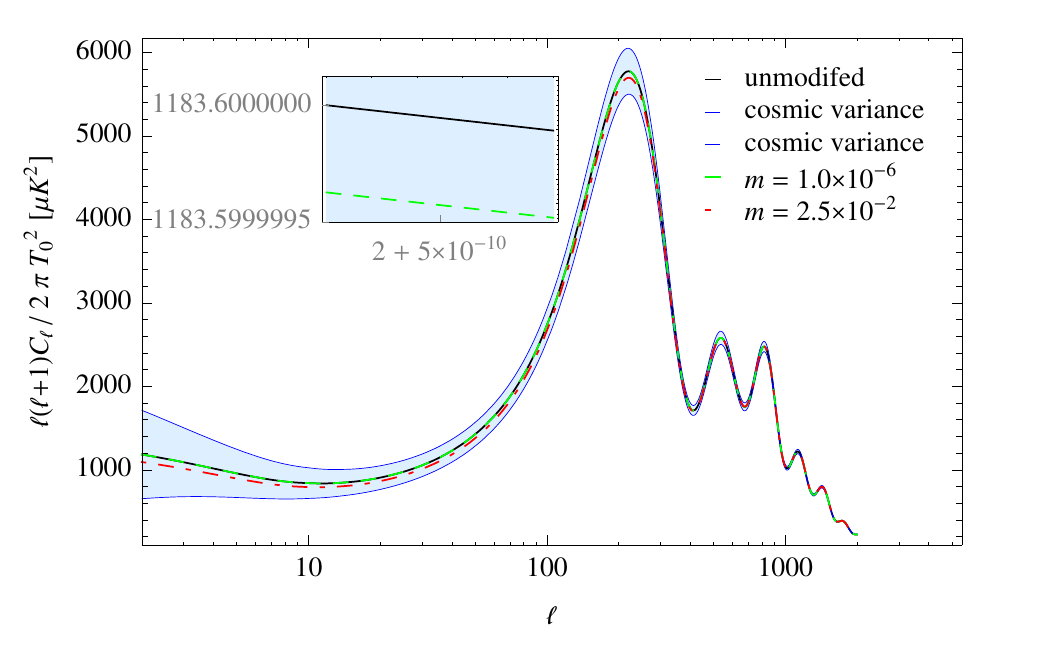}
	\vs{-5mm}
	\caption{Comparison of the power spectra $\ell (\ell + 1) \Crm_{\ell} / (2 \pi) T_{0}^{2}$ for the standard case
			(black, solid), as well as for the quantum-depletion corrected results ($m = 10^{-6}\.\mpl$, green, dashed;
			$m = 2.5 \cdot 10^{-2}\.\mpl$, red, dot-dashed). The shaded region depicts the cosmic-variance band.
			Masses are in units of $\mpl$, and the case $m = 2.5 \cdot 10^{-2}\.\mpl$ is included
			for illustrative purposes only.}
	\label{fig:Cl}
\end{figure}

In Fig.~\ref{fig:Cl}, we present the full power spectrum $\ell \,(\ell + 1) \,\Crm_{\ell} / (2 \pi)\,T_{0}^{2}$,
where $T_{0}$ is the CMB temperature today.
Here, this suppression is visible for small $\ell$, albeit it would only be notable
for rather large values of the inflaton mass.
To this end, and for illustrative purposes only, we again show the relative change to
the power spectrum for $m = 2.5 \cdot 10^{-2}\,\mpl$, bearing in mind that the
normalisation for this case would not match the Planck one.
Note that, even for this rather high value, all corrections remain well within the cosmic-variance
band.
Of course, the precise shape of this band is slightly model dependent.

\begin{figure}
	\centering
	\includegraphics[scale=1.40,angle=0]{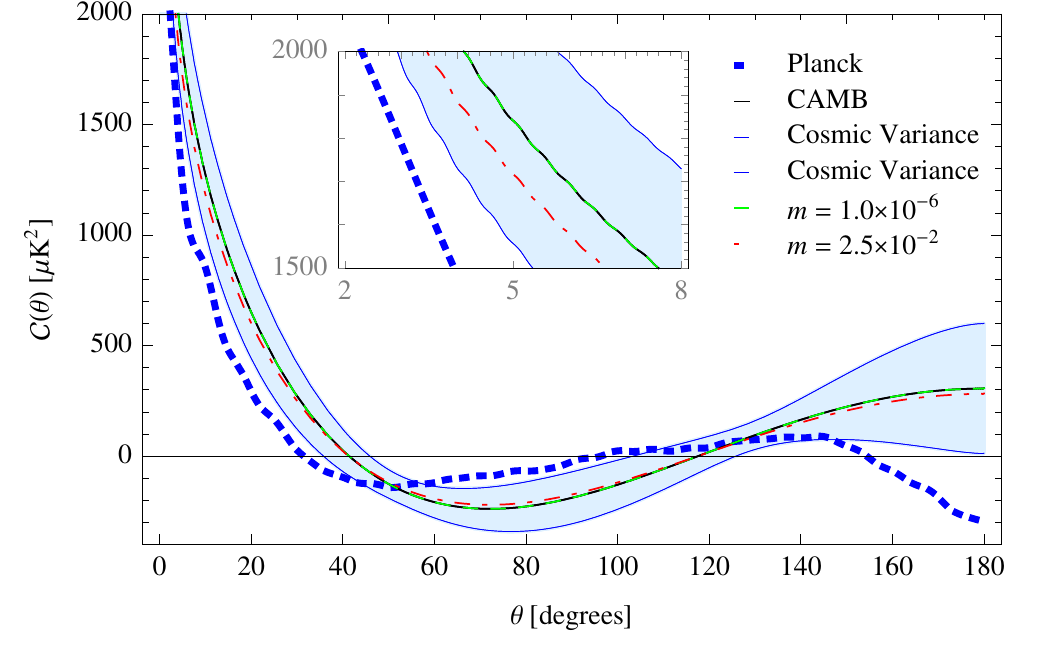}
	\vs{-2.9mm}
	\caption{Comparison of the angular temperature auto-correlation function $\Ccal( \theta )$ for the standard case
			(black, solid), the quantum-depletion corrected results ($m = 10^{-6}\.\mpl$, green, dashed;
			$m = 2.5 \cdot 10^{-2}\.\mpl$, red, dot-dashed), as well as for the Planck data (blue, thick, dotted).
			Masses are in units of $\mpl$, and the case $m = 2.5 \cdot 10^{-2}\.\mpl$ is included
			for illustrative purposes only.}
		\label{fig:CofTheta}
\end{figure}

Let us investigate whether this effect might also tell us something regarding
the seeming mismatch of the measured to the standard $\Lambda$CDM-calculated
angular temperature auto-correlation function $\Ccal( \theta )$ (\cf~Ref.~\cite{Copi:2013cya}
and references therein).
Therefore, in Fig.~\ref{fig:CofTheta}, we compare $\Ccal( \theta )$, which is computed
by summing multipole moments up to $\ell = 800$, for various cases (with and
without quantum-depletion modifications) to the Planck data.
We observe that the larger the mass, the more the corrected curves
approach the one inferred by the Planck measurements.
Although this effect seems too small to accommodate for the aforementioned
difference between measurements and the standard theory, we think that 
it deserves further investigations.

\section{Discussion and Outlook}
\label{sec:Discussion-and-Outlook}
\setcounter{equation}{0}

\noindent
In this work, we have quantitatively investigated the full time evolution
of inflationary space-time in the corpuscular description proposed by
Dvali and Gomez~\cite{Dvali:2013eja}.
We have computed several cosmological quantities, like the slow-roll
parameter $\epsilon$, the Hubble parameter $H$, and the tensor-to-scalar ratio $r$.
For all those quantities, we quantified the intrinsic quantum effects inherent
to the studied corpuscular structure of gravity.
These effects are actually stronger the longer inflation lasts,
and increase with increasing inflaton mass.

Regarding the tensor-to-scalar ratio $r$, we found that the quantum-depletion 
effects could be relevant, as $r$ is sensitive to the total duration
of the inflationary phase and these effects are cumulative in nature.
In particular, $r$ would be lower than in the standard semi-classical
computation for inflaton masses approaching the Planck scale.

We found that quantum depletion decreases the Hubble parameter,
whose amount we precisely quantified for different inflaton masses.
The gravitational potential is, in turn, also suppressed, and so are the CMB
multipole moments.
We computed the precise amount of this suppression for several values
of the inflaton mass, and found that it is strongest for low multiples, or equivalently,
on large scales.
The suppression is also stronger again for increasing inflaton mass,
and is completely consistent with observations, in particular 
for the CMB multipoles and the temperature auto-correlation function.

Having studied the simplest inflationary model, consisting of a single real
scalar field with a canonical kinetic term, as well as a mass term, we are tempted to
numerically investigate corpuscular effects on more refined models of inflation,
although we find it remarkable that already the simple set-up studied here,
where just the functional dependences in the master equation~\eqref{eq:N-dot}
are retained, is capable of reproducing the standard results.

On the qualitative level, we believe that our findings are generic, as the
occurrence of the ``quantum clock'', originating from the quantum depletion
(and working against the semi-classical one) is a necessity in the corpuscular
picture (cf.~Ref.~\cite{Dvali:2014gua}), and the calculation of the relevant $\Scal$-matrix
elements do not really depend on the specific choice of the inflaton potential
(cf.~Ref.~\cite{Isermann}).
The direction in which the corrections are going should always be the same,
and they should be stronger for lower multipoles (which correspond to earlier
times of horizon crossing).
However, on the quantitative level, different potentials, like the monomial ones
considered by the Planck collaboration~\cite{Planck15}, will lead to different
strengths of the corrections.
On a related note, but investigating the question of the occurrence of eternal
inflation in the corpuscular picture, it was shown in Ref.~\cite{Kuhnel:2015yka}
that monomial potentials lead to qualitatively similar effects, and these will be
stronger the steeper the potential.

Finally, it would be very interesting to further investigate the cumulativeness
of the quantum corpuscular effects, which seems likely to both tighten and loosen
constrains on inflationary and possibly modified-\./\.quantum-gravity models.\\[-9mm]

\acknowledgements
\vs{-4mm}R.~C.~and A.~O.~are supported by the INFN grant FLAG.
F.~K.~acknowledges supported from the Swedish Research Council (VR)
through the Oskar Klein Centre.
It is a pleasure to thank Michele~Cicoli, Gia~Dvali, Alessandro~Gruppuso,
Octavian~Micu, Edvard~M{\"o}rtsell, Douglas~Spolyar, and Nico~Wintergerst
for fruitful discussions.
Parts of this work are based on observations obtained with
Planck (http://www.esa.int/Planck), an ESA science mission with instruments
and contributions directly funded by ESA Member States, NASA, and Canada.

\appendix
\section{Effect of Non-Condensed Quanta}
\label{app:Effect-of-Non--Condensed-Quanta}
\setcounter{equation}{0}

\noindent
The effect of condensed graviton quanta is to generate an effective Hubble
expansion parameter $H\sim N^{-1/2}$ that satisfies the usual Friedmann equation.
Their occupation number $N$ is partly driven by the presence of the inflaton,
and by the depletion, which result in the master equation~\eqref{eq:N-dot}.
However, if we want to take into account the effect of the non-condensed quanta,
we should also add their energy density to the energy density of the ``ground state'',
like it was done for black holes in Ref.~\cite{Casadio:2014vja}.
In fact, the depletion causes the ground-state energy to decrease by reducing $N$,
but the quanta scattered off the ground state have higher energy, which should
affect the Hubble radius.
\par
Let us then replace $\rho$ with $\rho + \delta \rho$, where $\delta\rho$ is the
energy density of depleted gravitons.
This will of course affect the Hubble parameter $H^{2} \sim \rho$ and the slow-roll
parameter $\epsilon \simeq m^{2} / ( 3\.H^{2} )$.
The perturbation $\delta \rho$ is already provided in Ref.~\cite{Dvali:2013eja},
and is given by
\be
	\delta \rho = \frac{\delta N_\lambda}{\Rh^{3}} \frac{\mpl \lp}{\Rh}
	\ ,
\ee
where $N_\lambda$ is the number of deplete particles of wavelength $\lambda$,
both gravitons and inflatons.
From Eq.~\eqref{eq:gammagf} we can see that
$\delta N_{\phi} = \delta N = {N_{\phi}}/{N}$, which gives us
\be
	\delta \rho = \mpl \,\lp\, \frac{H}{\Rh^{3}}\,\frac{N_{\phi}}{N}
	\ .
\ee
Since the slow-roll parameter is also expressed via $\epsilon = - \dot{H} / H^{2}$,
we finally obtain the dynamical equation
\be
\dot{N}( t )
=
4\left(\frac{m}{\mpl}\right)^{5/2}\,
\frac{N^{9 / 4}( t )\, \sqrt{1 + m\,N^{3 / 2}( t )/\mpl} }
{ \lp\left(15 + 6\,m\,N^{3 / 2}( t )/\mpl\right) }
-
\frac{ \mpl }{ \lp\,m\,N( t )}
\ .
\label{eq:N-dot-full}
\ee
Note that Eq.~\eqref{eq:N-dot-full} precisely reduces to Eq.~\eqref{eq:N-dot}
for large $N$.
\par
\begin{figure}
	\centering
	\includegraphics[scale=1.40,angle=0]{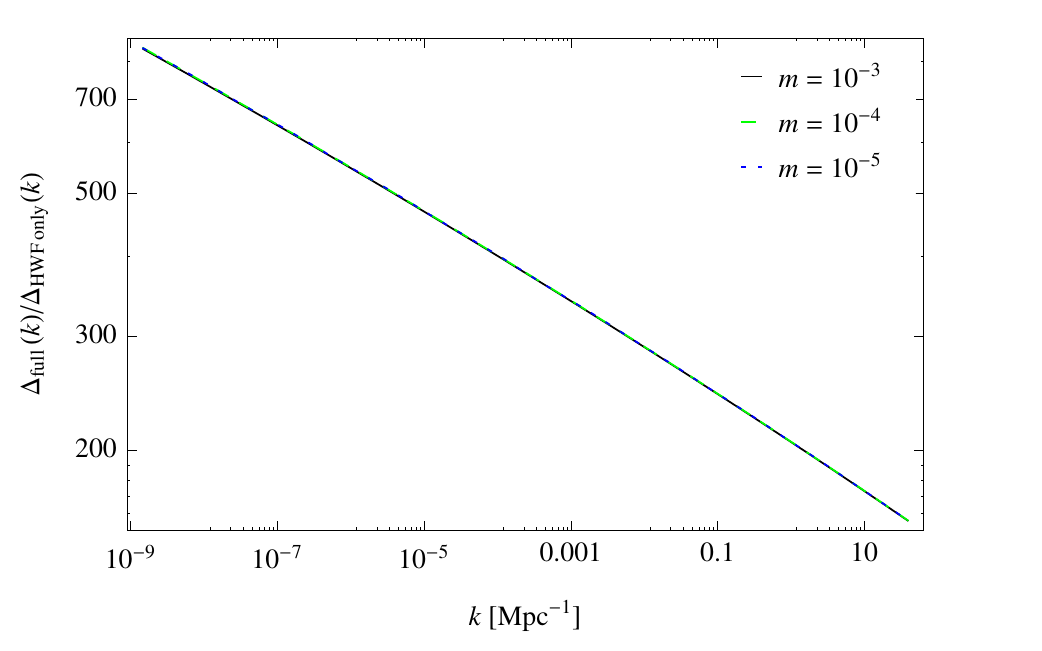}
	\caption{The ratio $\Delta_{\rm full} / \Delta_{\rm HWFonly}$ as a function of wave-number $k$;
			masses in the Figure are in units of $\mpl$.}
	\label{fig:HWFRelativeStrength}
\end{figure}

As before, we shall use the quantity
$\Delta( k ) \defas 1 - H_{\rm corrected}^{2}( k ) / H_{\rm uncorrected}^{2}( k )$
as a measure of the relative strength of a given correction.
In Fig.~\ref{fig:HWFRelativeStrength} we then display the ratio
$\Delta_{\rm full} / \Delta_{\rm HWFonly}$, where $\Delta_{\rm full}$ 
is computed from the evolution given by the full Eq.~\eqref{eq:N-dot-full}, whereas 
$\Delta_{\rm HWFonly}$ is obtained by subtracting the quantum-depletion term
from that equation.~\footnote{$\Delta_{\rm HWFonly}$ thus represents the correction
due to the higher energy of the excited quanta only, which would be obtained
by applying the formalism of the horizon wave function~\cite{HWF} to the present case.}
We observe that this ratio is essentially independent of the chosen inflaton
mass and is always much larger than one.
This shows that the correction omitted in the analysis of
Sec.~\ref{sec:Cosmological-Parameters} is indeed negligible,
and one can just employ the master equation
\be
\dot{N}( t )
=
\frac{ 2 }{ 3 }\.\frac{m^{2}\,N^{3 / 2}( t )}{\mpl^{2}\, \lp}
-
\frac{ \mpl }{ \lp\,m\,N( t ) }
\ .
\label{eq:N-dotA}
\ee

\begin{figure}
	\centering
	\includegraphics[scale=1.40,angle=0]{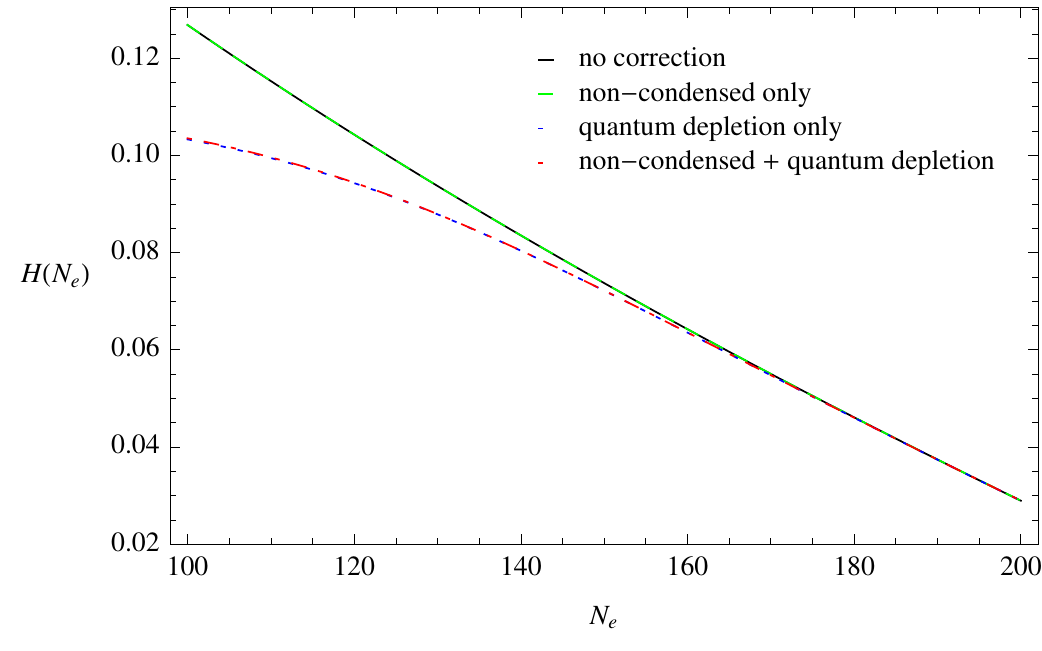}
	\caption{The Hubble parameter $H( N_{\erm} )$ as a function of the number of e-foldings
			for an inflaton mass $m = 10^{-2}\,\mpl$, for various quantum corrections.}
	\label{fig:HComparison}
\end{figure}

In Fig.~\ref{fig:HComparison} we illustrate the influence of the various quantum corrections
on the time-dependence of $H$, for various quantum corrections.~\footnote{A rather large inflaton
mass of $m = 10^{-2}\.\mpl$ is again used in order to clearly represent the influence
of those corrections.}

Let us finally note that the de~Sitter condition $\dot H = 0$, or, $\dot N = 0$
can be obtained for a number of graviton quanta in the BEC of order
\be
	N
		\sim 
								\bigg(
									\frac{ \mpl }{ m }
								\bigg)^{6 / 5}
								\ ,
\ee
which, for the typical inflaton mass $m\simeq 10^{13}\,$GeV, would be of order $10^{7}$.

\end{document}